\documentclass[11pt]{article}
\usepackage{amssymb}
\newcommand{\idty}{{\leavevmode{\rm 1\mkern -5.4mu I}}}
\newcommand{\be}{\begin{eqnarray}}
\newcommand{\ee}{\end{eqnarray}}
\newcommand{\A}{{\cal A}}

\renewcommand{\O}{\Omega}

\renewcommand{\d}{{\rm d}}
\newcommand{\D}{{\rm D}}
\newcommand{\bez}{\begin{eqnarray*}}
\newcommand{\eez}{\end{eqnarray*}}

\title{Bi-differential calculus and the KdV equation}

\author{ A. Dimakis \\ Department of Mathematics, University of the Aegean \\ 
        GR-83200 Karlovasi, Samos, Greece \\ dimakis@aegean.gr 
       \\[2ex]
         F. M\"uller-Hoissen \\ Max-Planck-Institut f\"ur Str\"omungsforschung \\
        Bunsenstrasse 10, D-37073 G\"ottingen, Germany \\
        fmuelle@gwdg.de }

\begin{document}
\renewcommand{\theequation} {\arabic{section}.\arabic{equation}}
\date{}
\maketitle
\begin{abstract}
A {\em gauged bi-differential calculus} over an associative (and not necessarily commutative) algebra $\A$ is an ${\mathbb{N}}_0$-graded left $\A$-module with two covariant derivatives acting on it which, as a consequence of certain (e.g., nonlinear differential) equations, are flat and anticommute. As a consequence, there is an iterative construction of generalized conserved currents. We associate a gauged bi-differential calculus with the Korteweg-de-Vries equation and use it to compute conserved densities of this equation.
\end{abstract}

\section{Introduction}
A distinguishing feature of soliton equations and other completely integrable models is the existence of an infinite set of conservation laws. For the special case of two-dimensional (principal) chiral or $\sigma$-models, a simple iterative construction of conserved currents and charges had been presented in \cite{BIZZ79}. In \cite{DMH96,DMH97,DMH98} some  generalizations of this work in the framework of noncommutative geometry have been achieved. In a recent work \cite{DMH99}, the existence of an infinite set of conserved currents in several completely integrable classical models, including chiral and Toda models, as well as the KP and self-dual Yang-Mills equations, has been traced back to a simple construction of an infinite chain of closed (respectively, covariantly constant) 1-forms in a (gauged) bi-differential calculus. A bi-differential calculus consists of a graded algebra on which two anticommuting differential maps act. In a gauged bi-differential calculus these maps are extended to covariant derivatives which, as a consequence of, e.g., nonlinear differential equations, are flat and anticommuting. 
\vskip.1cm

Section 2 introduces a mathematical scheme which may be regarded as the crucial structure behind the appearance of an infinite chain of conserved currents in the abovementioned completely integrable models (see also \cite{DMH99}). Section 3 shows how to realize such a scheme in terms of bi-differential calculi and covariant derivatives.
Section 4 treats the case of the Korteweg-de-Vries equation in some detail. Section 5 contains some conclusions.

\section{The central mathematical construction}
\label{sec:math}
\setcounter{equation}{0}
Let $\A$ be an associative algebra over $\mathbb{R}$ or $\mathbb{C}$ with a unit $\idty$. In the following, a {\em linear} map is meant to be linear over $\mathbb{R}$, respectively $\mathbb{C}$. We consider an ${\mathbb{N}}_0$-graded left $\A$-module ${\cal M} = \sum_{r \geq 0} {\cal M}^r$, on which two linear maps $\D, {\cal D} \, : \,  {\cal M}^r \rightarrow {\cal M}^{r+1}$ act such that
\be
   \D^2 = 0 \, , \quad  {\cal D}^2 =0 \, , \quad 
   {\cal D} \, \D = g \, \D \, {\cal D}           \label{bi-diff-cond}
\ee
with some $g \in \A$. Furthermore, we assume that, for some $s > 0$, there is a (nonvanishing) $\chi^{(0)} \in {\cal M}^{s-1}$ with 
\be
         {\cal D} \chi^{(0)} =0   \; .      \label{Dchi=0}
\ee
Then
\be
         J^{(1)} = \D \chi^{(0)} 
\ee
is ${\cal D}$-closed, i.e.,
\be
        {\cal D} J^{(1)} = g \, \D {\cal D} \chi^{(0)} = 0   \; .
\ee
If every ${\cal D}$-closed element of ${\cal M}^s$ is ${\cal D}$-exact, then
\be
       J^{(1)} = {\cal D} \chi^{(1)}
\ee
with some $\chi^{(1)} \in {\cal M}^{s-1}$. Now let $J^{(m)} \in {\cal M}^s$ satisfy
\be
       {\cal D} J^{(m)} = 0 \, , \qquad  J^{(m)} = \D \chi^{(m-1)}   \; .
\ee 
Then
\be
        J^{(m)} = {\cal D} \chi^{(m)}  
\ee
with some $\chi^{(m)} \in {\cal M}^{s-1}$ (which is determined only up to addition of some $\beta \in {\cal M}^{s-1}$ with ${\cal D} \beta=0$), and
\be
        J^{(m+1)} = \D \chi^{(m)}  
\ee
is also ${\cal D}$-closed:
\be
  {\cal D} J^{(m+1)} = g \, \D {\cal D} \chi^{(m)} = g \, \D J^{(m)} 
                     = g \, \D^2 \chi^{(m-1)} = 0  \; .
\ee
In this way one obtains an infinite tower of ${\cal D}$-closed elements $J^{(m)} \in {\cal M}^s$ and elements $\chi^{(m)} \in {\cal M}^{s-1}$ which satisfy
\be
      {\cal D} \chi^{(m+1)} = \D \chi^{(m)}  \; .   \label{D-chim-eq}
\ee
In certain cases this construction may break down at some level $m>0$ or become trivial in some sense (see also \cite{DMH99}). In terms of 
\be
  \chi = \sum_{m=0}^\infty \lambda^m \, \chi^{(m)}    \label{chi-sum}
\ee
with a parameter $\lambda$, the set of equations (\ref{D-chim-eq}) leads to
\be
    {\cal D} \chi = \lambda \, \D \, \chi  \; .    \label{D-chi-eq}
\ee
Conversely, if the last equation holds for all $\lambda$, we recover (\ref{D-chim-eq}).

\section{Bi-differential calculi and covariant derivatives}
\label{sec:bi-diff}
In this section we consider realizations of the structure introduced in the last section in terms of covariant exterior derivatives.

\vskip.1cm
\noindent
{\em Definition 1.}
A {\em graded algebra} over $\A$ is an ${\mathbb{N}}_0$-graded associative algebra $\O(\A) = \bigoplus_{r \geq 0} \O^r(\A)$ such that $\O^0(\A) = \A$ and the unit $\idty$ of $\A$ extends to a unit of $\O(\A)$, i.e., $\idty \, w = w \, \idty = w$ for all  $w \in \O(\A)$. 
\vskip.1cm
\noindent
{\em Definition 2.}
A {\em differential calculus} $(\O(\A),\d)$ over $\A$ consists of a graded algebra $\O(\A)$ over $\A$ and a linear map $ \d \, : \,  \O^r (\A) \rightarrow \O^{r+1}(\A)$ with the properties  
\be
       \d^2 &=& 0 \, ,  \label{d2=0}     \\
       \d (w \, w') &=& (\d w) \, w' + (-1)^r \, w \, \d w'  \label{Leibniz}
\ee
where $w \in \O^r(\A)$ and $w' \in \O (\A)$.\footnote{The identity $\idty \idty = \idty$ then implies $ \d \idty = 0 $.} We also require that $\d$ generates $\O(\A)$ in the sense that $\O^{r+1}(\A) = \A \, (\d \O^r(\A)) \, \A$. 
\vskip.1cm
\noindent
{\em Definition 3.}
A triple $(\O(\A),\d , \delta)$ consisting of a graded algebra $\O(\A)$ over $\A$ and two linear maps 
$\d , \delta \, : \, \O^r (\A) \rightarrow \O^{r+1}(\A)$ with the properties (\ref{d2=0}), (\ref{Leibniz}) and
\be
       \delta \, \d + \d \, \delta = 0   \label{d-delta}
\ee
is called a {\em bi-differential calculus}.
\vskip.1cm

Let $(\O(\A), \d , \delta)$ be a bi-differential calculus, and $A,B$ two $N \times N$-matrices of 1-forms  (i.e., the entries are elements of $\O^1(\A)$). We introduce
\be
     \D = \d + A \, \qquad  {\cal D} = \delta + B
\ee
which act from the left on $N \times M$-matrices with entries in $\O(\A)$. The latter form an ${\mathbb{N}}_0$-graded left $\A$-module ${\cal M} = \bigoplus_{r \geq 0} {\cal M}^r$. Then the conditions (\ref{bi-diff-cond}) with $g=-1$ can be expressed in terms of $A$ and $B$ as follows,
\be
      \D^2 = 0  \quad &\Longleftrightarrow&  \quad F = \d A + AA =0 \, , 
                \label{D2A}   \\
  {\cal D}^2 = 0  \quad &\Longleftrightarrow&  \quad {\cal F} = \delta B + BB =0  \, ,
                \label{D2B}   \\
  \D \, {\cal D} + {\cal D} \, \D = 0 \quad &\Longleftrightarrow& \quad  
      \d B + \delta A + BA + AB =0  \; .
                \label{DADB}
\ee
If these conditions are satisfied, we speak of a {\em gauged bi-differential calculus}.
\vskip.1cm

If $B=0$, the  conditions (\ref{D2A})-(\ref{DADB}) become $F =0$ and $\delta A =0$. There are two obvious ways to further reduce the latter equations: 
\vskip.1cm  
\noindent
(i) We can solve $F=0$ by setting $A = g^{-1} \, \d g$ with an invertible $N \times N$-matrix $g$ with entries in $\A$. The remaining equation reads $\delta (g^{-1} \, \d g) = 0$ which resembles the field equation of principal chiral models (cf \cite{DMH99}). 
\vskip.1cm
\noindent
(ii) We can solve $\delta A = 0$ via $A = \delta \phi$ with a matrix $\phi$. Then we are left with the equation
$\d (\delta \phi) + (\delta \phi)^2 = 0$ which generalizes the so-called `pseudodual chiral models' (cf \cite{Curt+Zach94} and references cited there).

\section{Example: conserved densities of the Korteweg-de-Vries equation}
\label{sec:KdV}
\setcounter{equation}{0}
Let $\A_0 = C^\infty({\mathbb{R}} \times {\cal I})$ be the algebra of smooth functions of coordinates $t, x$, where ${\cal I}$ is an interval, and $\A$ the noncommutative algebra generated by the elements of $\A_0$ and the partial derivative $\partial_x = \partial/\partial x$ such that $\partial_x f = f_x + f \, \partial_x$ for $f \in \A$. Here, $f_x$ denotes the partial derivative of $f$ with respect to $x$. Furthermore, let $\O^1(\A)$ be the $\A$-bimodule generated by two elements $\tau$ and $\xi$ which commute with all elements of $\A$. With 
\be
   \tau \, \xi = - \xi \, \tau \, , \quad 
   \tau \, \tau = 0 = \xi \, \xi
\ee 
we obtain a graded algebra $\O(\A) = \bigoplus_{r = 0}^2 \O^r(\A)$ over $\A$. Now
\be
  \d f &=& [ \partial_t + 4 \partial_x^3 , f ] \, \tau 
          - 6 \, [ \partial_x^2 , f ] \, \xi  \nonumber \\
       &=& ( f_t +4 f_{xxx} +12 \, f_{xx} \, \partial_x 
          +12 \, f_x \, \partial_x^2 ) \, \tau 
          - 6 \, ( f_{xx} + 2 \, f_x \, \partial_x ) \, \xi   \, ,    \\
  \delta f &=& - {1 \over 2} \, [ \partial_x^2 , f ] \, \tau 
             + [ \partial_x , f ] \, \xi 
           = - {1 \over 2} \, ( f_{xx} + 2 \, f_x \, \partial_x ) \, \tau + f_x \, \xi   \ee
and
\be
  \d (f \, \tau + h \, \xi) = (\d f) \, \tau + (\d h) \, \xi \, , \quad
  \delta (f \, \tau + h \, \xi) = (\delta f) \, \tau + (\delta h) \, \xi
\ee
define two linear maps $\d, \delta \, : \, \O^r(\A) \rightarrow \O^{r+1}(\A)$, and 
$(\O(\A),\d,\delta)$ becomes a bi-differential calculus  over $\A$. 
\vskip.1cm
\noindent
{\em Remark.} The above calculus is {\em noncommutative} in the sense that differentials do not, in general, commute with elements of $\A$, even with those of the commutative subalgebra $\A_0$. In particular, we have $x \, \delta x = (\delta x) \, x + \tau$. A (noncommutative) differential calculus is a basic structure in `noncommutative geometry'.  \rule{5pt}{5pt} 
\vskip.1cm

With $B=0$ and $A \in \O^1(\A)$, (\ref{DADB}) becomes $\delta A=0$ which is solved by 
\be
   A = \delta v = - {1 \over 2} ( v_{xx} + 2 \, v_x \, \partial_x ) \, \tau + v_x \, \xi
\ee
with $v \in \A_0$. Then $F=0$ takes the form
\be
    v_{tx} + v_{xxxx} - v_x \, v_{xx} = 0   \; .  
\ee
With the substitution
\be
      u = - v_x
\ee
this becomes the {\em Korteweg-de-Vries} equation
\be
    u_t + u_{xxx} + u \, u_x = 0   \; .          \label{KdV}
\ee

Let ${\cal M} = \O(\A)$. The general solution of $\delta \chi^{(0)} =0$ for 
$\chi^{(0)} \in \A$ is 
\be
   \chi^{(0)} = \sum_{n=0}^\infty c_n(t) \, \partial_x^n
\ee
with functions $c_n$ depending on $t$ only. A particular solution is given by 
$\chi^{(0)} = 1$. The equation (\ref{D-chi-eq}) with ${\cal D} = \delta$ is equivalent to the two equations
\be
    \chi_x &=& - \lambda \, ( 6 \, \chi_{xx} + u \, \chi 
               + 12 \, \chi_x \, \partial_x ) \, ,  \label{chix1}   \\
   - {1 \over 2} \, \chi_{xx} &=& \lambda \, \left( \chi_t + 4 \, \chi_{xxx} 
   + {1 \over 2} \, u_x \, \chi + u \, \chi_x + 6 \, \chi_{xx} \, \partial_x
     \right)  \; .   \label{chix2}
\ee
With 
\be
    \chi = \sum_{n=0}^\infty \chi_n \, \partial_x^n
\ee
the first equation is turned into the following set of equations,
\be
   \chi_{0,x} + \lambda \, ( 6 \, \chi_{0,xx} + u \, \chi_0 ) &=& 0  
                \label{chix01}  \\       
   \chi_{n,x} + \lambda \, ( 6 \, \chi_{n,xx} + 12 \, \chi_{n-1,x} 
     + u \, \chi_n ) &=& 0  \quad (n>0)  \; .
\ee
Inserting\footnote{Direct use of the expansion (\ref{chi-sum}) for $\chi_0$ leads to {\em nonlocal} conserved densities. The transformation from $\chi_0$ to $\varphi$ and subsequent expansion of $\varphi$ leads to {\em local} expressions, however.}
\be
    \chi_0 = e^{-\lambda \, \varphi}  \, , \qquad  
    \varphi = \sum_{m=0}^\infty (6 \lambda)^m \, \varphi^{(m)} 
\ee
(which sets $\chi^{(0)} = 1$) in (\ref{chix01}), we get
\be 
  \varphi_x = u - 6 \, \lambda \, \varphi_{xx} + 6 \, \lambda^2 \, (\varphi_x)^2
\ee
which in turn leads to
\be
  \varphi^{(0)}_x = u  \, ,  \qquad    \varphi^{(1)}_x =  - u_x
\ee
and
\be
   \varphi_x^{(m)} = - \varphi_{xx}^{(m-1)} + {1 \over 6} \, \sum_{k=0}^{m-2} \varphi_x^{(k)} \, \varphi_x^{(m-2-k)} 
\ee
for $m > 1$. Hence
\be
  \varphi^{(2)}_x &=& u_{xx} + {1 \over 6} \, u^2 \, ,          
                                                  \\
  \varphi^{(3)}_x &=& - ( u_{xx} + {1 \over 3} \, u^2 )_x   \, ,
                                                  \\
  \varphi^{(4)}_x &=& {1 \over 6} \, [ {1 \over 3} \, u^3 - (u_x)^2 ] + [ u_{xxx} 
                      + {1 \over 2} \, (u^2)_x ]_x   \, , 
                                                   \\
  \varphi^{(5)}_x &=& - [ {4 \over 27} \, u^3 + {5 \over 6} \, (u_x)^2 
                 + {4 \over 3} \, u \,u_{xx} + u_{xxxx} ]_x  \, ,
                                                    \\
  \varphi^{(6)}_x &=& {5 \over 216} \, [ u^4 - 12 \, u \, (u_x)^2 + {36 \over 5} \, (u_{xx})^2 ]                                                      \nonumber \\
                  & & + [ u_{xxxxx} + {5 \over 3} \, u \, u_{xxx} + {5 \over 6} \, u^2 \, u_x 
                      + 3 \, u_x \, u_{xx} ]_x   \, ,        
                                                     \\
  \varphi^{(7)}_x &=& - [ {2 \over 27} \, u^4 + {4 \over 3} \, u^2 \, u_{xx} 
    + {5 \over 3} \, u \, (u_x)^2 + {14 \over 3} \, u_x \, u_{xxx} + 2 \, u \, u_{xxxx} 
      \nonumber \\
                  & & + {10 \over 3} \, (u_{xx})^2  + u_{xxxxxx} ]_x   \, ,           
                                                    \\
  \varphi^{(8)}_x &=& {7 \over 648} \, [ u^5 - 30 \, u^2 (u_x)^2 + 36 \, u \, (u_{xx})^2 
                      - {108 \over 7} \, (u_{xxx})^2 ]  \nonumber \\
    & & + [ u_{xxxxxxx} + {7 \over 3} \, u \, u_{xxxxx} + {20 \over 3} \, u_x \, u_{xxxx} 
        + {35 \over 3} \,  u_{xx} \, u_{xxx}  \nonumber                  \\
    & & + {35 \over 18} \, u^2 \, u_{xxx} 
        + {95 \over 54} \, (u_x)^3 + {35 \over 216} \, (u^4)_x 
        + {7 \over 2} \, (u^2)_x u_{xx} ]_x  \, ,
                                                     \\
  \varphi^{(9)}_x &=& - [ {16 \over 405} \, u^5 + {20 \over 9} \, u^2 \, (u_x)^2 
        + {32 \over 27} \, u^3 \,  u_{xx} + {113 \over 9} \, (u_x)^2 \, u_{xx} 
        + {80 \over 9} \, u \, (u_{xx})^2           \nonumber \\
    & & + {112 \over 9} \, u \, u_x \, u_{xxx} + {8 \over 3} \, u^2 \, u_{xxxx} 
        + {23 \over 2} \, (u_{xxx})^2 + {56 \over 3} \,  u_{xx} \, u_{xxxx} 
                             \nonumber \\
    & & + 9 \, u_x \, u_{xxxxx} + {8 \over 3} \, u \, u_{xxxxxx} 
        + u_{xxxxxxxx} ]_x   \, , 
                                                      \\
  \varphi^{(10)}_x &=& {7 \over 1296} \, [ u^6 - 60 \, u^3 \, (u_x)^2 
        + 108 \, u^2 \, (u_{xx})^2 -30 \, (u_x)^4 - {648 \over 7} \, u \, (u_{xxx})^2 
                                                     \nonumber \\
    & & + {720 \over 7} \, (u_{xx})^3 + {216 \over 7} \, (u_{xxxx})^2 ]                                                     
        + [ u_{xxxxxxxxx} + {35 \over 72} \, u^4 \, u_x 
        + {35 \over 18} \, u^3 \, u_{xxx}  \nonumber \\
    & & + {21 \over 2} \, u^2 \, u_x \, u_{xx} 
        + {95 \over 18} \, u \, (u_x)^3
        + {7 \over 2} \, u^2 \, u_{xxxxx} + 20 \, u \, u_x \, u_{xxxx}           
                                           \nonumber \\
    & & + {455 \over 18} \, (u_x)^2 \,  u_{xxx} + 35 \, u \, u_{xx} \, u_{xxx} 
        + {69 \over 2} \, u_x \, (u_{xx})^2 
        + 3 \, u \, u_{xxxxxxx} 
                                           \nonumber \\
    & & + {35 \over 3} \, u_x \, u_{xxxxxx} + 28 \, u_{xx} \, u_{xxxxx} 
        + {125 \over 3} \, u_{xxx} \, u_{xxxx} ]_x  
\ee
and so forth. 
As a consequence of (\ref{chix1}) and (\ref{chix2}), we have 
\be
   \chi_{0,t} + \chi_{0,xxx} + {1 \over 2} \, u \, \chi_{0,x} = 0   \; . 
\ee
In terms of $\varphi$ this reads
\be
   \varphi_t + \varphi_{xxx} - 3 \, \lambda \, \varphi_x \, \varphi_{xx}
   + \lambda^2 (\varphi_x)^3 + u \, \varphi_x/2 = 0
\ee
and application of $\partial_x$ leads to a conservation law for $\varphi_x$,
\be
   \varphi_{xt} = -(\varphi_{xxx} - 3 \, \lambda \, \varphi_x \, \varphi_{xx}
   + \lambda^2 (\varphi_x)^3 + u \, \varphi_x/2 )_x  \; .
\ee
Hence, the $\varphi^{(m)}_x$ obtained above are conserved densities of the 
KdV equation. Let
\be
       Q^{(m)} = \int_{\cal I} \varphi^{(m)}_x \, dx  \label{Qm}
\ee
where $dx$ is the ordinary (Lebesgue) measure on $\mathbb{R}$. Here we assume that either $u$ is periodic in $x$ or that $u$ and its $x$-derivatives vanish sufficiently rapidly at the (finite or infinite) boundaries of the interval $\cal I$, so that the above integrals exist (see also \cite{MGK68}). Note that $\varphi^{(m)}$ will not, in general, be periodic or vanish at the ends of the interval, however. Now we have
\be
     {d \over dt} Q^{(m)} = \int_{\cal I} \varphi^{(m)}_{x t} \, dx = 0  \; .
\ee
Neglecting $x$-derivatives (which do not contribute to (\ref{Qm})) in the expressions for $\varphi_x^{(m)}$, we observe that $Q^{(m)} = 0$ for odd $m$. The nonvanishing conserved charges are
\be
     Q^{(0)} &=& \int_{\cal I} u \, dx    \\
     Q^{(2)} &=& {1 \over 6} \, \int_{\cal I} u^2 \, dx    \\
     Q^{(4)} &=& {1 \over 6} \, \int_{\cal I} [{1 \over 3} \, u^3 - (u_x)^2 ] \, dx  \\
     Q^{(6)} &=& {5 \over 216} \, \int_{\cal I} [ u^4 - 12 \, u \, (u_x)^2 + {36 \over 5} 
                 \, (u_{xx})^2 ] \, dx  \\
     Q^{(8)} &=& {7 \over 648} \, \int_{\cal I} [ u^5 - 30 \, u^2 \, (u_x)^2 + 36 \, u \, (u_{xx})^2 
                  - {108 \over 7} \, (u_{xxx})^2 ] \, dx  \qquad \\
     Q^{(10)} &=& {7 \over 1296} \, \int_{\cal I} [ u^6 - 60 \, u^3 \, (u_x)^2 
        + 108 \, u^2 \, (u_{xx})^2 -30 \, (u_x)^4   \nonumber \\
    & & - {648 \over 7} \, u \, (u_{xxx})^2 + {720 \over 7} \, (u_{xx})^3 
        + {216 \over 7} \, (u_{xxxx})^2 ]  \, dx   
\ee
and so forth. The integrands are in agreement (up to irrelevant constant factors) with $T_1, \ldots, T_6$ in  \cite{MGK68}, equations (5a)-(10a). Using computer algebra, it is easy to compute higher conserved charges. In \cite{KMG70} the uniqueness of the above sequence of conserved polynomial densities of the Korteweg-de-Vries equation has been shown. Therefore, the remaining freedom in the above construction cannot lead to additional polynomial conserved densities. 
\vskip.1cm
\noindent
{\em Remark.} Application of the central construction in section 2 to the case under consideration requires that $\delta$-closed elements of $\O^1(\A)$ are $\delta$-exact. $J \in \O^1(\A)$ can be written as $J = a \, \tau + b \, \xi$ with $a,b \in \A$. Then $\delta J =0$ means $a+b_x/2+b \, \partial_x=c$, where $c_x=0$. Introducing $\chi(t,x) = \int^x b(t,x') \, dx'$, we have $J = (c - {1 \over 2} \, \chi_{xx} -\chi_x \, \partial_x ) \, \tau + \chi_x \, \xi = \delta \chi + c \, \tau$. This does not work, however, for periodic boundary conditions on $\cal I$ (so that $\cal I$ is actually replaced by the circle $S^1$), since the indefinite integral of a periodic function $b$ need not be periodic. We still have the problem that the 1-form $\tau$ is not $\delta$-exact in $\O(\A)$. But with an extension of $\A$ and $\O(\A)$ (see also \cite{DMH99}, section 5.3) it becomes exact. This amounts to setting $\tau = \delta y$ with an additional coordinate $y$. Then $\delta$-closed elements of ${\cal M}^1$ are indeed $\delta$-exact. 
\rule{5pt}{5pt}

\section{Conclusions}
\label{sec:concl}
\setcounter{equation}{0}
The existence of a gauged bi-differential calculus as a (non-trivial) consequence of certain (e.g., differential, difference, or operator) equations may turn out to be a common feature of completely integrable systems. In \cite{DMH99} we have demonstrated that this concept covers many of the known soliton equations and other (in some sense) integrable models. The relation with various notions of complete integrability and approaches towards a classification of integrable models still has to be explored further. Moreover, the notion of a (gauged) bi-differential calculus and its generalization considered in section 2 applies to a large variety of structures (based on noncommutative algebras) most of which are far away from classical completely integrable models. It generalizes a characteristic feature of such models, namely the existence of an infinite set of conserved currents, into a framework of noncommutative geometry where an appropriate notion of complete integrability according to our knowledge is not yet at hand. 

\vskip.1cm
\noindent
{\em Acknowledgment.} F. M.-H. thanks the organizers of the XXXI Symposium on Mathematical Physics for the opportunity to lecture on related material in a splendid scientific and social atmosphere.

\end{document}